\documentstyle[12pt]{ioplppt}
\def \R{{\rm I\kern -0.13em R}}
\def \N{{\rm I\kern -0.13em N}}
\def \D{{\rm I\kern -0.13em D}}

\def \gam{\gamma}
\def \gams{\tilde{\gam}}

\def \V{\{V^\mu\}_\mu}
\newcommand{\vek}[1]{\mbox{\boldmath$#1$}}
\newcommand{\smvek}[1]{\mbox{\boldmath${\scriptstyle #1}$}}
\newcommand{\wkt}[1]{{\cal P}\left(#1\right)}
\newcommand{\ord}[1]{{\cal O}\left(#1\right)}

\newcommand{\beq}{\begin{equation}}
\newcommand{\eeq}{\end{equation}}
\newcommand{\beqa}{\begin{eqnarray}}
\newcommand{\eeqa}{\end{eqnarray}}
\newcommand{\beqas}{\begin{eqnarray*}}
\newcommand{\eeqas}{\end{eqnarray*}}

\begin{document}


\jl{1}
\title{Statistical Mechanics of Learning in the Presence of Outliers}
\author{Rainer Dietrich}
\address{Physikalisches Institut, Julius-Maximilians-Universit\"at,
Am Hubland, \\ D--97074 W\"urzburg, Federal Republic of Germany}
\author{Manfred Opper}
\address{Department of Computer Science and Applied Mathematics,
Aston University, \\ Birmingham B4\ 7ET, U.K.}
\begin{abstract}
Using methods of statistical mechanics, we analyse the effect of
outliers on the supervised learning of a classification problem.
The learning strategy aims at selecting informative examples
and discarding outliers. We compare two algorithms which
perform the selection either in a soft or a hard way.
When the fraction of outliers grows large, the estimation
errors undergo a first order phase transition.
\end{abstract}
\pacs{87.10.+e, 05.90.+m}
\maketitle

\section{ Introduction }

The analysis of algorithms which allow to learn a rule from random examples
is an active and fascinating topic in the area of statistical mechanics.
For an overview see e.g. \cite{Seu91,Waal93,OpKi}.
Many models, where examples are {\em correctly} classified
by ideal experts (often called teachers) seem to be well understood.
Now, there is a great deal of interest in nonideal, but more realistic
models, which incorporate the influence of different types of noise
in learning.

In this paper, we study a model where not all examples
carry information about the unknown rule, but where a nonzero
fraction of them are just outliers.
Naively learning {\em all} examples may considerably
deteriorate the ability to infer the rule in such a case.
Similar to learning with noisy data, some knowledge about the stochastic
data generating mechanism can be helpful. Based on such a stochastic
model, a good algorithm could try to select the informative examples
and discard the remaining ones. Since however only partial information is
available, such a selection can only be performed approximately and
it is natural to try a {\em soft}, probabilistic selection.

Our model leads naturally to such a selection method. It consists of
a classification problem, where data which come from two
distributions (classes) centered at different points
are mixed at random with outliers.
A Bayesian approach, which aims at calculating the most probable
values for the class centers by minimizing a specific {\em training
energy} is combined with the
so-called EM algorithm of Dempster et al \cite{DeLaRu}, which nicely deals
with the problem of hidden parameters (the knowledge which of the data
are informative) in data mixtures.
This procedure leads to an algorithm which iteratively computes
the probability that an example is informative and
weights each example in predicting the unknown
class centers of the data generating distributions.
Our model may also be considered as a simple version of the
{\em mixtures of experts} models
\cite{JaJoHi} which are frequently studied
in the neural network literature. In these models, a complicated
task is learnt by a division of labor among several simple learning
machines (experts), where each
expert learns from different subsets of examples. Our model would
correspond to two experts where only one is able to extract
information from the examples.

The paper is organized as follows: After an introduction of the learning
problem, two learning strategies are defined in section two.
Section three gives the statistical mechanics
formulation of the problem, which, based on a replica calculation,
leads to a computation of the learning performance in the
thermodynamic limit. In section four the algorithmic implementation
of the learning methods using the EM algorithm is explained.
Section five presents the results of the statistical mechanics
calculations and of numerical simulations and concludes with a discussion.
Details of the replica calculations are given in the appendices.

\section{The Learning Problem}

We assume that the examples $\{\vek{\xi}^\mu,S^\mu\}$
($\vek{\xi}^\mu \in \R^N, S^\mu \in \{\pm1\}$),
$\mu=1,\ldots,\alpha N$, are generated alternatively
by two different processes. For the first process, the
input $\vek{\xi}^\mu$
is selected at random from one of two gaussian
clusters (labelled by the outputs $S^\mu=\pm 1$) which are chosen
with equal probability. The clusters are centered
at $\pm\vek{B}$ and have equal variance $1/\gam$.
$\vek{B}$ is an $N$ dimensional vector with $\vek{B}^2/N=1$.
The joint probability for inputs and outputs corresponding to this
process can be written as
$$
\wkt{\vek{\xi}^\mu,S^\mu |\vek{B}} \propto
        \exp \left[ -\frac{\gam}{2} \sum_j
\left( \xi_j^\mu - \frac{1}{\sqrt{N}} S^\mu B_j \right)^2 \right].\\
$$
The data from this process represent classified examples in a
noisy (because the Gaussian clusters overlap) two-class problem.

In the second process, the inputs come from a single gaussian centered
at zero with the same variance and the output
(chosen $\pm 1$ with equal probability) is
completely independent from the input.
For this case, we make the ansatz
$$
\wkt{\vek{\xi}^\mu,S^\mu |\vek{B}} \propto
        \exp \left[ -\frac{\gam}{2} \sum_j \left( \xi_j^\mu \right)^2 \right].
$$
The data from the second process may be understood as representing
outliers which do not contain any information about the two spatially
structured classes of inputs and
come from a "garbage" class and are classified purely by random guessing.
In order to distinguish the two processes, we introduce decision
variables $V^\mu\in\{0,1\}$, where $V^\mu=1$ stands
for the first process and $V^\mu=0$ for the outliers. The joint set of
decision variables is denoted by $\{V^\mu\}_\mu$. Conditioning on
these variables, we can write the probability distribution for the
the joint set of $\alpha N$ data
$ \D := \{ \vek{\xi}^\mu,S^\mu \}_\mu $,
$ \mu = 1,\ldots,p=\alpha N $ within the single equation
\beqa\label{distexam}
\wkt{\D \mid \{V^\mu\}_\mu, \vek{B}} = \\ \nonumber
\frac{1}{2^{\alpha N}} \left( \frac{\gam}{2\pi} \right) ^{\alpha N^2 /2}
        \prod_{\mu,j} \exp \left[ -\frac{\gam}{2} (\xi_j^\mu)^2
        + \frac{\gam}{\sqrt{N}} V^\mu \xi_j^\mu S^\mu B_j
        - \frac{\gam}{2N} V^\mu {B_j}^2 \right].
\eeqa
In order to model the fact that outliers occur at random with a fixed rate,
we will assume that both processes (structure, outliers)
are chosen independently at random. The probability for having
the value $V^\mu$ is written as
\beqa\label{distv}
\wkt{V^\mu} &= & \frac{ \exp[-\eta V^\mu] }{ 1 + \exp[-\eta] }.
\eeqa
Using the "chemical potential" $\eta$, we can adjust the average fraction
of structured data
\beqas
\overline{V^\mu} & = & \frac{ 1 }{ \exp[\eta] + 1 }.
\eeqas
For $\eta=-\infty$ all examples have $V^\mu=1$, but
with increasing $\eta$, less examples carry information.
For $\eta=0$, only half of the examples come from the structure
and for $\eta=\infty$ all examples are outliers.

A learner tries to infer the vector $\vek{B}$
from the $\alpha N$ examples and makes an estimate
$\vek{J}$ for $\vek{B}$.
We will assume that the fraction of outliers
is known to the learner. Although in our final results we will
mostly deal with the case that also the
parameter $\gam$ is known precisely, we will be more general in the
basic definitions and assume that the learner
uses $\gams$ instead, with $\gam \neq \gams$.
Hence, if the $\{V^\mu\}_\mu$ were known, the likelihood of the data
based on the estimate $\vek{J}$ would be given by
\beqas
\wkt{\D \mid \{V^\mu\}_\mu, \vek{J}} = \\
        \frac{1}{2^{\alpha N}} \left( \frac{\gams}{2\pi} \right) ^{\alpha
		N^2 /2}
        \prod_{\mu,j} \exp \left[ -\frac{\gams}{2} (\xi_j^\mu)^2
        + \frac{\gams}{\sqrt{N}} V^\mu \xi_j^\mu S^\mu J_j
        - \frac{\gams}{2N} V^\mu {J_j}^2 \right].
\eeqas
In general, however, the learner does not know which of the examples
contain information and which are outliers. Hence, to the learner the
$\{V^\mu\}_\mu$ are
{\em hidden variables} which are not observed but need to be averaged over.
Hence, the actual ansatz for the distribution of data will be given by
the {\em mixture distribution}
\beq\label{mix}
\wkt{\D \mid \vek{J}} = \sum_{\{V^\mu\}_\mu}
\wkt{\D, \{V^\mu\}_\mu \mid \vek{J}},
\eeq
where
\beqa\nonumber
\wkt{\D, \{V^\mu\}_\mu \mid \vek{J}}
        = \wkt{\D \mid \{V^\mu\}_\mu, \vek{J}} \wkt{\{V^\mu\}_\mu} \\
        = \frac{1}{2^{\alpha N}} \left( \frac{\gams}{2\pi} \right) ^{\alpha
		N^2 /2}
              \frac{ 1 }{ ( 1 + \exp[-\eta] )^{\alpha N} }
              \exp \left[ -\frac{\gams}{2} \sum_{\mu,j} (\xi_j^\mu)^2
              - \sum_\mu V^\mu f_\mu(\vek{J}) \right]
\eeqa
and where we have defined
$$
f_\mu(\vek{J})
        := -\frac{\gams}{\sqrt{N}} \sum_j \xi_j^\mu S^\mu J_j
              + \frac{\gams}{2N} \sum_j {J_j}^2 + \eta.
$$
One possible way of getting an estimate for the unknown vector
$\vek{B}$, would be the {\em maximum likelihood} method, i.e., one would
use the vector $\vek{J}$ which maximizes the likelihood (\ref{mix}).
A second possibility is given by a Bayesian approach, where the learner
supplies some {\em prior knowledge} about reasonable estimates
$\vek{J}$ within a {\em prior distribution}. We will use a distribution
which on average gives the correct length of the unknown vector
but does not favour any spatial direction
\begin{equation}\label{prior}
\wkt{\vek{J}} = \left( \frac{ 1 }{ 2\pi } \right)^{N/2}
        \exp \left[ -\frac{1}{2} \sum_j {J_j}^2 \right].
\end{equation}
Based on the prior and the likelihood of the data, the learner can construct
the posterior distribution, using Bayes rule
\begin{equation}\label{poster}
\wkt{\vek{J} \mid \D} =
\frac{\wkt{\D \mid \vek{J}}  \wkt{\vek{J}}}{\wkt{\D}}.
\end{equation}
There are several ways of using the information contained in the posterior
(\ref{poster}). E.g., simply taking the {\em posterior mean} as the estimate
for $\vek{B}$ will minimize the expected average
(with respect to the posterior)
squared error. Unfortunately, for a high dimensional space, such
expectations will not be easy to calculate exactly, and one has to resort
to Monte Carlo sampling. A simpler estimate, which should not perform too
poorly, is given by the vector $\vek{J}$, which has maximal aposteriori
probability (MAP), i.e., the one which maximizes (\ref{poster}).
Actually, if there are enough data available, one can expect that
the posterior will be
close to a gaussian, and both estimates will come close.

In order to maximize the posterior
$\wkt{\vek{J} \mid \D}$ with respect to $\vek{J}$, we can equivalently
minimize the "training"-energy function
\beqa\label{energy}
{\cal H}(\vek{J})=-\ln \wkt{\D, \vek{J}}
& = &
-\ln \sum_{\{V^\mu\}_\mu} \wkt{\D, \{V^\mu\}_\mu \mid \vek{J}} \wkt{\vek{J}}.
\eeqa
As we will see in section four, there is a simple algorithm to calculate the
MAP. As we will see, this algorithm is based on a recursive estimation
of the (posterior) expected decision variables $\{V^\mu\}_\mu$.
Since examples will be weighted by their probability of being informative
rather than being kept or discarded from the training set, we call this
method a {\em soft selection} of examples.

As an alternative to the MAP approach for $\vek{J}$,
we will discuss also an algorithm which calculates the MAP for the
hidden variables $\{V^\mu\}_\mu$. Since these variables take the
values $0$ and $1$ only, the result will be a {\em hard} selection of
informative examples, rather than a soft weighting.
We look for the values of $\{V^\mu\}_\mu$
which maximize
\beq
\wkt{\{V^\mu\}_\mu \mid \D} = \frac{\wkt{\D , \{V^\mu\}_\mu}}
{\wkt{\D}}.
\eeq
Equivalently, we can maximize the numerator of this expression, which can
be written as a mixture probability
\beq\label{mix:two}
\wkt{\D , \{V^\mu\}_\mu} = \int d \vek{J}\, \wkt{\D , \{V^\mu\}_\mu, \vek{J}}
\eeq
resulting in a training energy
\beq\label{henergy}
{\cal H}_h(\V) = - \ln \int d\vek{J} \, \wkt{\D,\vek{J},\V}.
\eeq
Finally, after minimization, we can use the expectations
\beq
{\left\langle J_j \right\rangle}_{\smvek{J}}
= \frac{\int d\vek{J} \, J_j \, \wkt{\D,\vek{J},\V}}
{\int d\vek{J} \, \wkt{\D,\vek{J},\V}}
\eeq
as an estimate for the unknown $B_j$.

\section{ Analysis by Statistical Mechanics }

In this section, we study the performance of both
MAP estimates analytically in the thermodynamic limit $N\to\infty$
using a statistical mechanics framework.
We begin first with the soft selection.
There are different ways of measuring, how good the learner, equipped
with the MAP estimate, has learnt the structured distribution.
An obvious idea is to measure the quadratic deviation between
the true vector $\vek{B}$ and the MAP:
\begin{equation}
\Delta = \frac{1}{N}
\left\langle \left( \vek{J}-\vek{B} \right)^2 \right\rangle = Q - 2 R + 1
\end{equation}
where we have defined the order parameters
\begin{eqnarray}\nonumber
R = \frac{1}{N} \left\langle \vek{J} \cdot \vek{B} \right\rangle  \\
Q = \frac{1}{N} \left\langle \vek{J} \right\rangle ^2.
\end{eqnarray}
It is also useful to calculate the
angle $\Phi=\angle (\vek{J},\vek{B})$ between estimate and $\vek{B}$.
This angle $\Phi$, normalized by $1/\pi$ is given in terms of the
order parameters by
\beqa\label{phi}
\Phi & = & \frac{1}{\pi} \arccos \frac{\vek{J}\cdot\vek{B}}
{||\vek{J}||\ ||\vek{B} ||}\\
& = & \frac{1}{\pi} \arccos \frac{R}{\sqrt{Q}}
\eeqa
The order parameters for the soft selection MAP algorithm
can be derived from a partition function
$Z$ where the corresponding hamiltonian is given by
${\cal H}(\vek{J})$ from (\ref{energy}).
Assuming that the inverse temperature $\beta$ is an integer, we define
\begin{eqnarray}\nonumber
Z & = & \int d\vek{J} \, \exp \left[ -\beta {\cal H}(\vek{J})  \right] \\
\nonumber
& = & \int d\vek{J} \, \exp \left[ \beta \ln \wkt{\D, \vek{J}} \right] \\
\nonumber
& = & \int d\vek{J} \, \left(\wkt{\D, \vek{J}}\right)^{\beta}\\
& = &  \int d\vek{J} \, \left\{ \sum_{\{V^\mu\}_\mu}
\wkt{\D, \{V^\mu\}_\mu,\vek{J}} \right\}^{\beta}\\
& = &  \int d\vek{J} \,
\sum_{\{V_{b}^\mu\}_\mu}
\prod_{b=1}^{\beta}\wkt{\D, \{V_b^\mu\}_\mu,\vek{J}}. \nonumber
\end{eqnarray}
The MAP, which is the minimum of
the energy ${\cal H}(\vek{J})$, is derived from the limit $\beta\to\infty$.
The case $\beta=1$ would correspond to
Gibbs learning, where a vector $\vek{J}$ is drawn at
random from the posterior.
As usual, order parameters are found from an average of the
free energy $ f = -\frac{1}{\beta N} \ln Z $ over the distribution
of the examples. To perform the average, we utilize the replica trick
\beqa\nonumber
\left\langle f \right\rangle & = & -\frac{1}{\beta N} \left\langle \ln Z
\right\rangle \\
        & = & -\frac{1}{\beta N} \lim_{n\to0}
\frac{\partial}{\partial n} \ln \left\langle Z^n \right\rangle
\eeqa
where $\langle \ldots \rangle$ denotes the average over
the distribution (see (\ref{distexam}) and (\ref{distv}))
$$
\wkt{\xi_j^\mu, S^\mu \mid \vek{B}} = \frac{1}{2} \left( \frac{\gam}{2\pi}
\right) ^{1/2}
        \frac{1}{1+e^{-\eta}}
        \sum_{V^\mu} \exp \left[ -\frac{\gam}{2} \left( \xi_j^\mu -
\frac{1}{\sqrt{N}} V^\mu S^\mu
        B_j \right) ^2 - \eta V^\mu \right].
$$
The replicated partition function is now written as
\beq
Z^n=\sum_{\{V_{ab}^\mu\}_\mu} \int \prod_{a} \, d\vek{J}^a \,
\prod_{a,b} \wkt{\D, \{V_{ab}^\mu\}_\mu,\vek{J}^a}.
\eeq
where the decision variables contain {\em two replica} indices.
Here, the index $a$ runs from $1$ to $n$, whereas $b$ runs from $1$
to $\beta$. For the subsequent calculations we have 
assumed the correct parameters $\gam=\gams$ and have
made a {\em replica symmetric ansatz} with respect to the indices $a$.
We think that this should be at least a good approximation,
because our model is an example of a 
{\em teacher--student} learning scenario, where student and teacher match
in the sense that the student uses the right statistical model for the data. 
For the Gibbs learning scenario ($\beta=1$), where the symmetry of 
student and teacher becomes perfect in the replica calculation
(this can be seen by introducing a further
average over $\vek{B}$, using the prior (\ref{prior})),
replica symmetry is usually considered to be exact
(however no general proof has been given sofar). 
Hence, assuming that the effects of replica symmetry breaking are
small, we have refrained  
from performing a replica stability analysis.

The treatment of the replica indices $b$ is much simpler, because the 
order parameters (see Appendix A) do not depend on them. 
Hence, as long as $\beta$ is an integer, no further symmetry
assumptions are required for the $b$'s. 
Although we don't have a proof that the continuation to
noninteger $\beta$ is unique, we expect that
the limit $\beta\to\infty$ exists and can be safely calculated 
using a sequence of integers.

The {\em hard selection problem} of decision variables is treated similarly
using the (zero temperature) free energy which is defined from the
partition function
\beq\label{zhard}
Z_h= \sum_{\{V^\mu\}_\mu} e^{-\beta {\cal H}_h(\V)}
\eeq
with the energy (\ref{henergy}).
The averages which are necessary for the calculation of error measures, e.g.
\beq
\Phi = \frac{1}{\pi} \arccos
	\frac{ \sum_j {\left\langle J_j \right\rangle}_{\smvek{J}} B_j }
	{ \sqrt{ \sum_j {\left\langle {J_j}^2 \right\rangle}_{\smvek{J}} }
	\sqrt{N}}
\eeq
can be found in a standard way from derivatives of the free energy with respect
to appropriate external fields, e.g.
\beqa\label{ordhard}
\sum_j {\left\langle J_j \right\rangle}_{\smvek{J}} B_j =
-\lim_{\lambda\to 0}\frac{\partial}{\partial\lambda}
\lim_{\beta\to\infty}\frac{1}{\beta}
\ln \sum_{\{V^\mu\}_\mu} e^{-\beta {\cal H}_h(\V,\lambda)}
\eeqa
where
\beqas
{\cal H}_h(\V,\lambda)=
-\ln \int d\vek{J} \, \wkt{\D,\vek{J},\V}
		\exp \left[ - \lambda \sum_j J_j B_j \right].
\eeqas
Explicit calculations of the free energies and order parameters
for both cases are given in the appendices.

\section{ The EM-Algorithm }

Unfortunately, the maximization of the posterior
distributions cannot be carried out in closed form
and must be done numerically. Usually, nonlinear optimization problems
are solved by gradient descent algorithms which require a tuning of the
step sizes. However, for the type of (generalized) maximum likelihood problem
for mixture distributions such as (\ref{mix}) and (\ref{mix:two}),
there is a simpler
and well known algorithm which has been developped by Dempster et al
\cite{DeLaRu}.
This so-called {\em expectation maximization (EM) algorithm}
guarantees that the (generalized) likelihood is nondecreasing for every
iteration step and converges to a local maximum.
To explain the idea for the soft selection problem, let us assume for
the moment that the hidden variables
$\{V^\mu\}_\mu$ were actually known. Then the corresponding log-likelihood
$\ln \left[\wkt{\D, \{V^\mu\}_\mu \mid \vek{J}} \wkt{\vek{J}}\right]$
could be maximized in closed form. In the EM algorithm, the true values of
the hidden variables are replaced iteratively by suitable averages.
At iteration $i$, in the {\em expectation step},
the function
\begin{equation}\label{expec}
A(\vek{J},\vek{J}^{(i)})
:= { \left\langle \ln \left[ \wkt{\D, \{V^\mu\}_\mu \mid \vek{J}}
\wkt{\vek{J}} \right] \right\rangle }_{ \wkt{\{V^\mu\}_\mu \mid \D ,
\vek{\scriptstyle J}^{(i)}} }
\end{equation}
is calculated, which is the log likelihood of observed and
hidden data averaged over the
posterior distribution of the hidden data, given the old estimate
$\vek{J}^{(i)}$. In the {\em maximization step}, (\ref{expec})
is maximized with respect to $\vek{J}$ in order to obtain the
new iteration $\vek{J}^{(i+1)}$.

We will not give the proof of convergence here,
as it is relatively simple and can be found in many
textbooks (see e.g. \cite{Honer}).
However, we can easily see that a fixed point of the algorithm is
also a local extremum of (\ref{energy}). At the maximum of (\ref{expec}),
we have
\beqas
0=\frac{\partial}{\partial J_k} A(\vek{J},\vek{J}^{(i)})
& = & \frac{\partial}{\partial J_k} {
\left\langle \ln \left[ \wkt{\D, \{V^\mu\}_\mu \mid \vek{J}}
\wkt{\vek{J}} \right] \right\rangle }_{ \wkt{\{V^\mu\}_\mu \mid \D ,
\vek{\scriptstyle J}^{(i)}}} \\
& = & \sum_{\{V^\mu\}_\mu}\frac{\frac{\partial}{\partial J_k}
\wkt{\D, \{V^\mu\}_\mu, \vek{J}}
\wkt{\D, \{V^\mu\}_\mu , \vek{J}^{(i)}}}
{\wkt{\D, \{V^\mu\}_\mu, \vek{J}}
\wkt{\D, \vek{J}^{(i)}}}.
\eeqas
Hence, at the fixed point, where $\vek{J}^{(i)}=\vek{J}$, we also have
$\frac{\partial}{\partial J_k}\ln \wkt{\D, \vek{J}} =0$.
For the explicit calculation, we need the
conditional
distribution of the hidden variables, given the data and $\vek{J}$
\beqa\label{postv}\nonumber
\wkt{\{V^\mu\}_\mu \mid \D , \vek{J}}
& = &
\frac{\wkt{\D, \{V^\mu\}_\mu , \vek{J}}}
{\wkt{\D, \vek{J}}} \\
& = &
\prod_{\mu} \frac{\exp[- V^\mu f_\mu(\vek{J})]}
{1+ \exp[- f_\mu(\vek{J})]}.
\eeqa
Using the distribution (\ref{distv}), we get
\beqas
\frac{\partial}{\partial J_k} A(\vek{J},\vek{J}^{(i)})
& = & - \gams \sum_\mu \langle V^\mu \rangle \left( -\frac{1}{\sqrt{N}}
\xi_k^\mu S^\mu + \frac{1}{N} J_k \right) - J_k \\
& \stackrel{!}{=} & 0
\eeqas
which gives
\beq
\vek{J}  =  \frac{ \sqrt{N} \sum_\mu \langle V^\mu \rangle \vek{\xi}^\mu
S^\mu }
{ \sum_\mu \langle V^\mu \rangle + N/\gams },
\eeq
where
\beqa\nonumber
\langle V^\mu \rangle & = & \sum_{V^\mu=0,1} V^\mu
\wkt{V^\mu \mid \D,\vek{J}^{(i)}} \\
& = & \frac{1}{ \exp \left[ f_\mu (\vek{J}^{(i)}) \right] + 1 }.
\eeqa
Hence, the estimate $\vek{J}$ for $\vek{B}$
is of the form of a {\em weighted Hebbian} sum, where each example
has a weight which is proportional to the estimated probability
$\langle V^\mu \rangle$, that the example is not an outlier.
It is interesting to look at the limiting case $\eta\to -\infty$, i.e. where
all examples are from the double cluster and where no
outliers are present. In this case, the EM iteration stops after one step,
and we get
\begin{eqnarray}\nonumber
\langle V^\mu \rangle & = & 1 \mbox{ for all } \mu \\ \label{Hebb}
\vek{J} & = & \frac{1}{\sqrt{N}}
\frac{ \sum_\mu \vek{\xi}^\mu S^\mu }{ \alpha + 1/\gams }
\end{eqnarray}
which is the usual Hebbian vector.

Similarly, to apply the EM algorithm to
the hard selection problem with the mixture distribution (\ref{mix:two}),
we take $\vek{J}$ as the hidden quantity. In each iteration step,
we have to maximize
\begin{eqnarray}\nonumber
\hat{A}(\{V^\mu\}_\mu,\{V^\mu\}_\mu^{(i)}) & :
= & { \left\langle \ln \wkt{\D, \{V^\mu\}_\mu, \vek{J}} \right\rangle }
_{ \wkt{\vek{\scriptstyle J} \mid \D , \{V^\mu\}_\mu^{(i)}} } \\ \nonumber
        & = &   - \frac{\gams}{2} \sum_{\mu,j} (\xi_j^\mu)^2
                + \frac{\gams}{\sqrt{N}} \sum_{\mu,j} V^\mu \xi_j^\mu S^\mu
{\left\langle J_j \right\rangle} \\
        &   &   - \frac{\gams}{2N} \sum_{\mu,j} V^\mu {\left\langle {J_j}^2
\right\rangle}
                - \eta \sum_\mu V^\mu
                - \frac{1}{2} \sum_j {\left\langle {J_j}^2 \right\rangle}
\end{eqnarray}
with respect to $\{V^\mu\}_\mu$.
Defining
\begin{eqnarray}\nonumber
a   & := & \frac{\gams}{N} \sum_\mu V^{\mu\,(i)} + 1 \\
b_j & := & \frac{\gams}{\sqrt{N}} \sum_\mu V^{\mu\,(i)} \xi_j^\mu S^\mu
\end{eqnarray}
we obtain for the expectations at step $i$
\begin{eqnarray}\nonumber
{\left\langle J_j \right\rangle} & = & \frac{b_j}{a} \\
 & = & \frac{ \sqrt{N} \sum_\mu V^{\mu\,(i)} \xi_j^\mu S^\mu }{ \sum_\mu
V^{\mu\,(i)} + N/\gams } \\ \nonumber
{\left\langle {J_j}^2 \right\rangle} & = & \frac{{b_j}^2}{a^2} + \frac{1}{a}.
\end{eqnarray}
Finally, after convergence, we use ${\left\langle J_j \right\rangle}$ as
an estimate for $B_j$.

\section{Results and Discussion}

\subsection{Soft Selection}
Solving for the order parameters and assuming that  $\gams =\gam $
we find that for fixed $\eta$, as expected, both error measures $\Phi$
and $\Delta$ decrease
towards $0$ with an increasing number $\alpha N$ of examples, showing that
the algorithm is able to find the true structure vector $\vek{B}$.
Since for the EM algorithm both error measures show qualitatively the same
behaviour, we will concentrate mainly on the angle $\Phi$.

Fig.~1 shows $\Delta(\alpha)$ for $\eta=0$. The second curve
gives the performance of the Hebbian rule (\ref{Hebb}). It
demonstrates the importance of selecting informative examples.
If all examples are weighted equally (and $\eta\neq\-\infty$), then
the true vector $\vek{B}$ cannot be recovered for $\alpha\to\infty$.
In Fig.~2, $\Phi(\alpha)$ (EM algorithm) is shown for
$\eta=0$ and $\eta=4$.
Since it was harder to perform simulations for $\eta=4$,
where only about $1.8\%$ of the examples are informative, we have shown
simulations only for $\eta=0$.
Asymptotically one finds a decrease of the error like
\beq
\Phi
\stackrel{\alpha\to\infty}{\simeq} \frac{1}{\pi R_{\infty}}
\sqrt{\frac{c}{\alpha}},
\eeq
where $R_{\infty}$ is the asymptotic value of the orderparameter $R$
and both $R_{\infty}$ and $c$ depend on $\eta$.

As expected, for fixed $\alpha$, the error increases
with $\eta$, i.e. with a growing number of outliers. More interesting is
the nonsmooth behaviour of the second curve, which gives a sudden drop
of the error as $\eta$ is varied. This phase transition can be
observed in more detail in the relief plot of the order parameters
$R$ and $Q$ in Figs.~3a and 3b.
In regions of large $\eta$ or large $\alpha$, the saddlepoint
equations have three solutions. Taking the solution with the smallest
free energy leads to a jump of the order parameters.
It is easier to investigate the transition by simulations as a
function of $\eta$, for fixed $\alpha$. This is shown in Fig.~4,
together with the predictions of the theory.

We have simulated the EM-algorithm starting from random initial
conditions and averaged the order parameters over many samples
of random inputs. Fixing $\alpha$,
the simulations show a good agreement with the
theory for small and large values of $\eta$, but
discrepancies show up close to the predicted transition. 
Since the average fraction $\bar{V}$ of informative data points 
decreases exponentially with $\eta$, finite size effects
play a crucial role in the simulations. 
E.g.~for $\eta=4$, less than 2 examples
out of $N=100$ are informative on average whereas the replica 
theory is based on infinitely
many examples from the structured clusters. 
Hence, we have performed a finite size scaling to determine 
the critical value 
$\bar{V}_0$, where the transition sets in. Since for small 
$\eta$ (large $\bar{V}$), 
the simulations show rather small statistical fluctuations around 
a value of $R$ close to 1, 
we have (for each $N$) defined $\bar{V}_0$ as the point, where
the distribution of the observed values for $R$ significantly broadens,
indicating the onset of transitions to different values of $R$.
A simple linear extrapolation to $N=\infty$ as shown in the inset of 
Fig.~4 gives a value for $\bar{V}_0$ which is 
in good agreement with the predicted value for the phase transition.
The large error bar at $\eta=6.8$ is explained from the fact that
the values for $\Phi$ (eq.~(\ref{phi})) have been obtained by 
using the sample averages of $R$ and $Q$ which (for finite $N$)
show a transition at slightly different values of $\eta$.

\subsection{Hard Selection}

Solving the orderparameter equations for the free energy (\ref{freehard})
at zero temperature, we find similar first order transitions as
for the method of soft selection.
For $\eta$ small enough, there is only one solution
which has a nonzero overlap to the teacher vector $\vek{B}$.
Increasing $\eta$ (and thereby the
number of outliers) beyond a value $\eta_0$, another solution with
${\hat{R}}={\hat{Q}}={\hat{z}}=0$ (see eq.~(\ref{hardord}))
appears, i.e.~where all $V^{\mu}=0$
and all data are considered to be outliers. Here
\begin{equation}
\eta_0 = -\frac{\gams}{2} + \frac{\gams^2}{4\gam}.
\end{equation}
Between $\eta_0$ and a second parameter value $\eta_c$, however, this trivial
solution has a higher free energy $f_h=0$ than the nontrivial one.
Finally, for $\eta>\eta_c$, the trivial solution with zero
order parameters, giving rise to $\Phi=1/2$,
is the one with lowest free energy.
Fig.~5 shows this critical $\eta$ as a function of $\alpha$.

So, unlike in the soft selection case, we have, for a large range of $\eta$,
two solutions of the orderparameter equations. This is reflected in the
simulations, the single runs clearly tending to either of these two optima.
Effects of metastability (which would be a sign of a rugged energy
landscape and indicate strong effects of replica symmetry breaking) 
could not be observed. However, a finite size scaling 
for the transition point did not lead to a satisfactory agreement with the 
theory. We think that the observed discrepancy is a dynamical effect,
where the EM algorithm, starting from a random initial condition,
is unable to reach the global minimum and converges only to the local one, 
thus shifting the phase transition to smaller values of $\eta$. 
We have balanced this effect to some
extent by keeping only those simulations (as long as they
occur) where the EM algorithm converges to the solution with 
nonzero overlap to the vector $\vek{B}$.

Fig.~6 shows the performance of the hard selection for $\alpha=20$.
Comparision to Fig.~4 suggests that the soft selection should be preferred.
The difference between the performance of the two algorithms becomes
more drastic for $\alpha\to\infty$:
The soft selection algorithm is able to tolerate an {\em arbitrary fraction}
of outliers as long as enough data available. Eventually, it will always
find the true teacher vector $\vek{B}$. On the other hand, for hard selection,
the explicit solution of the orderparameter equations for
$\alpha\to\infty$ shows that there is always a critical fraction of outliers
(corresponding to a parameter $\eta_c$ (\ref{etacinf})),
where learning is no longer possible even
if inifinitely many examples are available.
It is also interesting to investigate the influence of the overlap
of the two gaussian clouds in the structured input distribution
on the transition parameter $\eta_c$.
Fig.~7 shows $\eta_c$ for $\alpha=\infty$ as a function of
$\gamma$, which gives the inverse squared width of each gaussian
and measures so the distinguishability of the clouds.
If $\gamma$ is below 0.278, somewhat surprising, the critical $\eta$
jumps discontinuously to zero, i.e. if the overlap of the two clouds
is above a certain value, only $50\%$ outliers can be tolerated.

Phase transitions in the performance of learning algorithms have been observed
frequently in the statistical mechanics of neural networks. Since
such effects do not occur in asymptotic (in the sense of large $\alpha$)
expansions or in the exact bounds known in statistics they seem to be
one of the major contributions of statistical mechanics to the field of
computational learning theory. Phase transitions occur in
multilayer networks, where they are can be related to
the breaking of symmetries which are related to the network
architecture \cite{SH,Op94}.
Other examples include models with a so-called student teacher
mismatch \cite{Gyoer2},
models with discrete adjustable parameters \cite{GarDer89,Gyoer90}
and models of unsupervised learning \cite{BiMi93,Bark}.
For the present supervised learning model, where the basic
adjustable parameters are continuous variables and where the learner
matches with the distribution of the data, the phase transition was
unexpected. It will be interesting to apply recently developped
combinations of statistical mechanics techniques and methods of
information theory \cite{OpHa}
to establish the existence of phase transitions in mixture models
in more general circumstances.

\begin{appendix}
\section{Free energy and order parameters for soft selection}

Upon averaging, we obtain
\beqas
\left\langle Z^n \right\rangle = \\
\sum_{\{V_{ab}^\mu\}_\mu}
\int \prod_{a,j} \,dJ_j^a
        \exp \left[ -\sum_{a,b} \left( \frac{\gams}{2N}
\sum_\mu V_{ab}^\mu
        + \frac{1}{2} \right) \sum_j (J_j^a)^2 -
\eta \sum_{a,b} \sum_\mu
        V_{ab}^\mu \right] \\ \nonumber
\times\left\langle \exp
\left[ -\frac{\tilde{\gam} n \beta}{2} \sum_{\mu,j} (\xi_j^\mu)^2
        + \frac{\gams}{\sqrt{N}}
\sum_{a,b} \sum_{\mu,j} V_{ab}^\mu \xi_j^\mu S^\mu
        J_j^a \right] \right\rangle.
\eeqas
Within replica symmetry, the introduction of the order parameters
\beqas
R = \frac{1}{N} \left\langle \vek{J} \cdot \vek{B} \right\rangle & =
& \frac{1}{N} \sum_j J_j^a B_j \\
q = \frac{1}{N} \left\langle \vek{J}^2 \right\rangle & =
& \frac{1}{N} \sum_j J_j^a J_j^{\tilde{a}} \\
Q = \frac{1}{N} \left\langle \vek{J} \right\rangle ^2 & =
& \frac{1}{N} \sum_j (J_j^a)^2
\eeqas
together with their conjugates yields
\beqas
\left\langle Z^n \right\rangle & \propto &
         \int \prod_{a,j} \,dJ_j^a
        \exp \left[ iN\Phi \left( \frac{1}{N} \sum_{j,a} J_j^a B_j - n R
\right) \right] \\
    & & \prod_{} \exp \left[ iN\omega \left( \frac{1}{N} \sum_{j,a,\tilde{a}
\neq a} J_j^a J_j^{\tilde{a}} - n(n-1)q \right) \right] \\
    & & \exp \left[ iN\Omega \left( \frac{1}{N} \sum_{j,a} (J_j^a)^2 - nQ
\right) \right]  \\
    & & \sum_{\{V_{ab}^\mu\}_\mu}\left( \prod_{a,b} \exp \left[ - \left(
\frac{\gams}{2} \sum_\mu V_{ab}^\mu + \frac{1}{2} N \right) Q - \eta \sum_\mu
V_{ab}^\mu \right] \right) \\
    & & \left( \prod_\mu \exp \left[ \frac{1}{1+n\beta\gams/\gam} \left(
-\frac{1}{2} \gams n \beta (V^\mu)^2 + \gams \sum_{a,b} V_{ab}^\mu
V^\mu R \right.\right.\right. \\
    & & \left.\left.\left. + \frac{\gams^2}{2\gam} \sum_{a,\tilde{a}\neq a}
\sum_{b,\tilde{b}} V_{ab}^\mu V_{\tilde{a}\tilde{b}}^\mu q
        + \frac{\gams^2}{2\gam} \sum_{a} \sum_{b,\tilde{b}} V_{ab}^\mu
V_{a\tilde{b}}^\mu Q \right) - \eta V^\mu \right] \right)
\eeqas
In this expression (and in the following one) the order parameters have to
be taken at their saddle point values.
After a lengthy calculation, we arrive at an expression for the free energy
\begin{equation}\label{freebeta}
f = \frac{1}{\beta} \frac{R^2-Q}{2(Q-q)} - \frac{1}{2\beta} \ln (Q-q)
+ \frac{1}{2} Q - \frac{\alpha}{\beta} M(R,q,Q) + \mbox{const.}
\end{equation}
with
\beqas
M(R,q,Q) = \frac{1}{1+e^{-\eta}} \int Dx \left\{ \ln \left( \int Dy
\left( 1 + \exp \left[ -\frac{\gams}{2} Q - \eta + \gams
\sqrt{\frac{q}{\gam}}x \right.\right.\right.\right. \\
\left.\left.\left. + \gams \sqrt{\frac{Q-q}{\gam}}y \right] \right) ^\beta
\right)
-\frac{1}{2} e^{-\eta} \gams\rho^2\beta \\
\left. + e^{-\eta} \ln \left( \int Dy \left( 1 + \exp \left[ -\frac{\gams}{2}
Q - \eta + \gams R + \gams \sqrt{\frac{q}{\gam}}x
+ \gams \sqrt{\frac{Q-q}{\gam}}y \right] \right) ^\beta \right) \right\}.
\eeqas
For $\beta\to\infty$ we have to take the limit $q\to Q$.
With the ansatz $(Q-q)\beta =: z = \ord{1}$, we get in the limit
\beq
f = \frac{R^2-Q}{2z} + \frac{1}{2} Q
- \frac{\alpha}{1+e^{-\eta}}
\left( \hat{I}_5 + \frac{b}{2} \hat{I}_1 \right)
        - \frac{\alpha e^{-\eta}}{1+e^{-\eta}} \left( I_5 + \frac{b}{2} I_1
\right) + \mbox{const.}
\eeq
This yields the saddlepoint equations
\beqas
0 \stackrel{!}{=} \frac{\partial f}{\partial R} & = &
        \frac{R}{z} - \frac{\alpha e^{-\eta}}{1+e^{-\eta}}
\left( I_6 + \frac{b}{2} I_2 \right) \gams \\
0 \stackrel{!}{=} \frac{\partial f}{\partial z} & = &
        \frac{Q-R^2}{2z^2} - \frac{\alpha}{1+e^{-\eta}} \hat{I}_4
\frac{\gams^2}{2\gam}
        - \frac{\alpha e^{-\eta}}{1+e^{-\eta}} I_4 \frac{\gams^2}{2\gam} \\
0 \stackrel{!}{=} \frac{\partial f}{\partial Q} & = &
        -\frac{1}{2z} + \frac{1}{2} - \frac{\alpha}{1+e^{-\eta}}
\left( \frac{\gams}{2} \left( \hat{I}_6 + \frac{b}{2} \hat{I}_2 \right)
+ \frac{\gams}{2\sqrt{\gam Q}} \left( \hat{I}_7 + \frac{b}{2} \hat{I}_3
\right) \right) \\
 & &    - \frac{\alpha e^{-\eta}}{1+e^{-\eta}} \left( \frac{\gams}{2}
\left( I_6 + \frac{b}{2} I_2 \right) + \frac{\gams}{2\sqrt{\gam Q}}
\left( I_7 + \frac{b}{2} I_3 \right) \right)
\eeqas
where
\beqas
I_1 & := & \int Dx \frac{1}{e^{-2a}+1+(2-b)e^{-a}} \\
I_2 & := & \int Dx \frac{2e^{-2a}+(2-b)e^{-a}}{\left( e^{-2a}+1+(2-b)e^{-a}
\right)^2} \\
I_3 & := & \int Dx \frac{2e^{-2a}+(2-b)e^{-a}}{\left( e^{-2a}+1+(2-b)e^{-a}
\right)^2} x \\
I_4 & := & \int Dx \frac{e^{-2a}+1+2e^{-a}}{\left( e^{-2a}+1+(2-b)e^{-a}
\right)^2} \\
I_5 & := & \int Dx \ln \left( 1+e^a \right) \\
I_6 & := & \int Dx \frac{1}{e^{-a}+1} \\
I_7 & := & \int Dx \frac{x}{e^{-a}+1}
\eeqas
For the $\hat{I}_j$, $a$ has to be replaced by $\hat{a}$, where
\beqas
a & := & -\frac{\gams}{2} Q - \eta + \gams R + \gams \sqrt{\frac{Q}{\gam}}x \\
\hat{a} & := & -\frac{\gams}{2} Q - \eta + \gams \sqrt{\frac{Q}{\gam}}x \\
b & := & \frac{\gams^2}{\gam} z.
\eeqas
\section{ Free energy and order parameters for hard selection }
The Hamiltonian (\ref{henergy}) is explicitely given by
\beqas
{\cal H}_h(\V) & := & - \ln \int d\vek{J} \, \wkt{\D,\vek{J},\V} \\
       & = & -\left[ \frac{\gams^2}{2 N (\gams \hat{Q} + 1)}
		 \sum_{\mu,\nu} V^\mu V^\nu \sum_j\xi_j^\mu \xi_j^\nu S^\mu
S^\nu
		\right. \\
	& &	\left.
		- \frac{\gams}{2} \sum_{\mu,j} (\xi_j^\mu)^2 -
                  \eta \sum_\mu V^\mu \right]
	     + (N/2)\ln(\gams\hat{Q} + 1) -\ln C
\eeqas
where
\beqas
C := \frac{1}{2^{\alpha N}} \left( \frac{\gams}{2\pi} \right) ^{\alpha N^2 /2}
		\frac{ 1 }{ ( 1 + \exp[-\eta] )^{\alpha N} }
		\left( \frac{1}{ 2\pi } \right)^{N/2}
\eeqas
with the orderparameters
\beqas\label{hardord}
{\hat{R}} & := & \frac{1}{N} \sum_\mu V_a^\mu V^\mu \\
{\hat{q}} & := & \frac{1}{N} \sum_\mu V_a^\mu V_{\tilde{a}}^\mu \\
{\hat{Q}} & := & \frac{1}{N} \sum_\mu (V_a^\mu)^2 = \frac{1}{N} \sum_\mu
V_a^\mu .
\eeqas
Averaging the partition function (\ref{zhard}) yields
\beqas
\left\langle Z_h^n \right\rangle & =
& 		\left( \frac{1}{1+e^{-\eta}} \right) ^{\alpha N}
		\left( \frac{1}{1+n \beta \gams/\gam} \right) ^{\alpha
N^2 / 2}
\sum_{\{V_a^\mu,V^\mu\}_\mu} \int \prod_{a,j} Dy_j^a
\\
	& &	\exp \! \left[ -\eta \sum_\mu V^\mu
    		+ \frac{1}{2(1+n \beta \gams/\gam)} \left(
    		- n \beta \gams \sum_\mu V^\mu
		\right. \right. \\
	& &	\left. \left.
    		+ \frac{2 \gams \sqrt{\beta}}{\sqrt{\gams \hat{Q} + 1}}
    		\sum_j \sum_a y_j^a B_j \hat{R}
    		+ \frac{\gams^2 \beta}{\gam (\gams \hat{Q} + 1)}
    		\sum_j \sum_{a,\tilde{a}\neq a} y_j^a y_j^{\tilde{a}} \hat{q}
		\right. \right. \\
	& &	\left. \left.
    		+ \frac{\gams^2 \beta}{\gam (\gams\hat{Q} + 1)}
    		\sum_j \sum_a (y_j^a)^2 Q \right)
    		- n N \beta \eta \hat{Q} \right]
		\\
	& &    (\gams Q + 1) ^{-n N \beta/2} C^{n \beta}
\eeqas
The free energy $f_h$ simplifies in the limit $\beta\to\infty$,
where the scaling
$
\beta ({\hat{q}}-{\hat{Q}}) =: {\hat{z}} = \ord{1}
$
is used.
We finally obtain $f_h$ as a function of the actual orderparameters at the
saddlepoint:
\beq\label{freehard}
f_h= \eta {\hat{Q}} + \frac{1}{2} \ln ({\hat{Q}}\gams+1)
		- \frac{({\hat{Q}}+2{\hat{R}}^2\gam\rho^2)({\hat{Q}}\gams+1)
\gam\gams^2}
		{4({\hat{Q}}\gam\gams+{\hat{z}}\gams^2+\gam)^2}
\eeq
A similar calculation using (\ref{ordhard}) yields the averages
\beqa\nonumber
\sum_j {\left\langle J_j \right\rangle}_{\smvek{J}} B_j
& = &	N \frac{ {\hat{R}} \gams }{ {\hat{Q}} \gams +1 }
	\left( 1 - \frac{ {\hat{z}}\gams^2 }
{ {\hat{Q}}\gam\gams+{\hat{z}}\gams^2+\gam } \right) \\
\sum_j {\left\langle {J_j}^2 \right\rangle}_{\smvek{J}}
& = &	N \frac{ \gam\gams^2 ({\hat{Q}}+2{\hat{R}}^2\gam) }
	{ 2({\hat{Q}}\gam\gams+{\hat{z}}\gams^2+\gam)^2 } +
N \frac{ 1 }{ {\hat{Q}}\gams+1 }.\label{ordhard:two}
\eeqa
In the limit $\alpha\to\infty$, the resulting order parameter equations
can be further simplified by making the scaling ans\"atze
${\hat{R}} = \alpha {\hat{R}}_0$,
${\hat{Q}} = \alpha {\hat{Q}}_0$,
${\hat{z}} = -\alpha {\hat{z}}_0$,
where ${\hat{R}}_0,{\hat{Q}}_0,{\hat{z}}_0$ are independent of
$\alpha$ as $\alpha\to\infty$.
For $\gam=\gams$, the equation for the critical ratio of
outliers $\eta_c$, where the trivial solution with zero orderparameters
has the global minimum of the free energy, is determined from
\beqa\label{etacinf}
0 & = & \eta - 2 \gam \pi \eta \exp[\gam+2\eta]
	\Phi^2[\sqrt{\gam}-\sqrt{2\eta}] /
	\left\{ \exp[\sqrt{2\gam\eta}] + \exp[\gam/2+\eta]
		\right. \\ \nonumber
	& &	\left.
	+ \sqrt{\pi\eta} \exp[\gam/2+\eta]
	\left( -2\Phi[\sqrt{\gam}-\sqrt{2\eta}] - 2 e^\eta +
2 e^\eta \Phi[\sqrt{2\eta}] \right) \right\} ^2.
\eeqa

\end{appendix}

\newpage


\newpage
\begin{figure}[ht]
\begin{center}
\setlength{\unitlength}{1mm}
\begin{picture}(150,100)
\put(0,0){\makebox(150,100)
          {\includegraphics{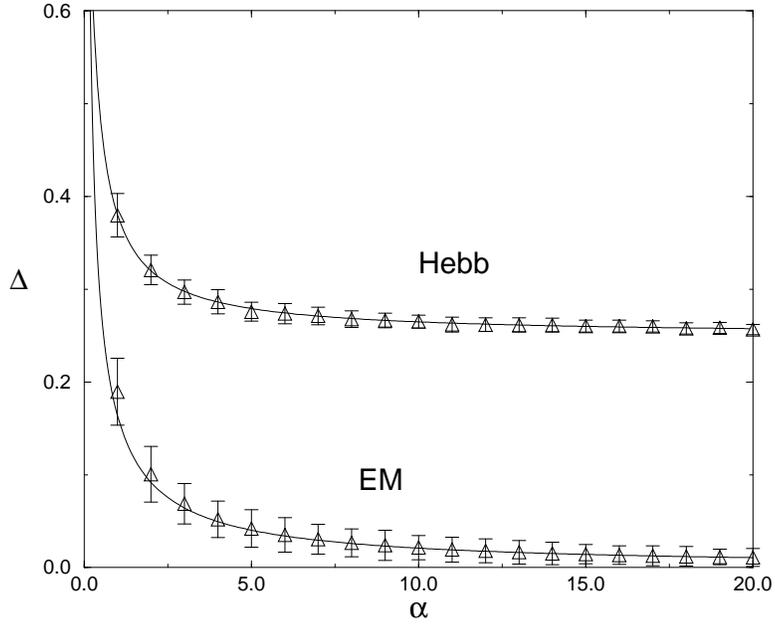}}}
\end{picture}
\end{center}
\caption{ 
Comparison between EM-algorithm and naive Hebb rule. Parameters are 
$\eta=0, \gam=\gams=10$. The solid lines show the theoretical
results. Simulations are done with $N=500$; here as in subsequent plots,
bars mark standard deviations over 100 runs.
 }
\end{figure}
\begin{figure}[ht]
\begin{center}
\setlength{\unitlength}{1mm}
\begin{picture}(150,100)
\put(0,0){\makebox(150,100)
          {\includegraphics{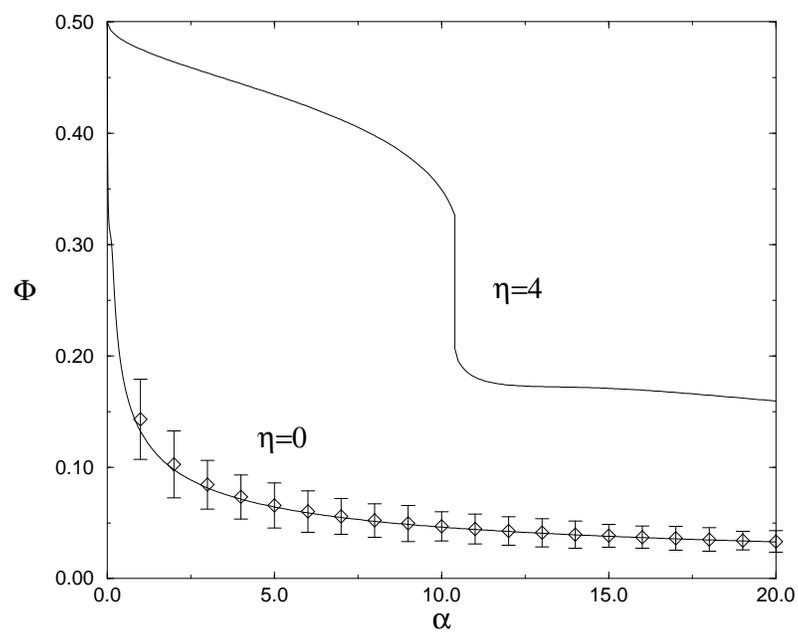}}}
\end{picture}
\end{center}
\caption{ 
$\Phi(\alpha)$ for $\eta=0$ and $\eta=4$, respectively (MAP estimate).
The simulations at $\eta=0$ are performed with $N=500$; results are 
averaged over 100 runs.  
 }
\end{figure}
\begin{figure}[ht]
\begin{center}
\setlength{\unitlength}{1mm}
\begin{picture}(160,200)
\put(0,100){\makebox(150,100)
          {\includegraphics{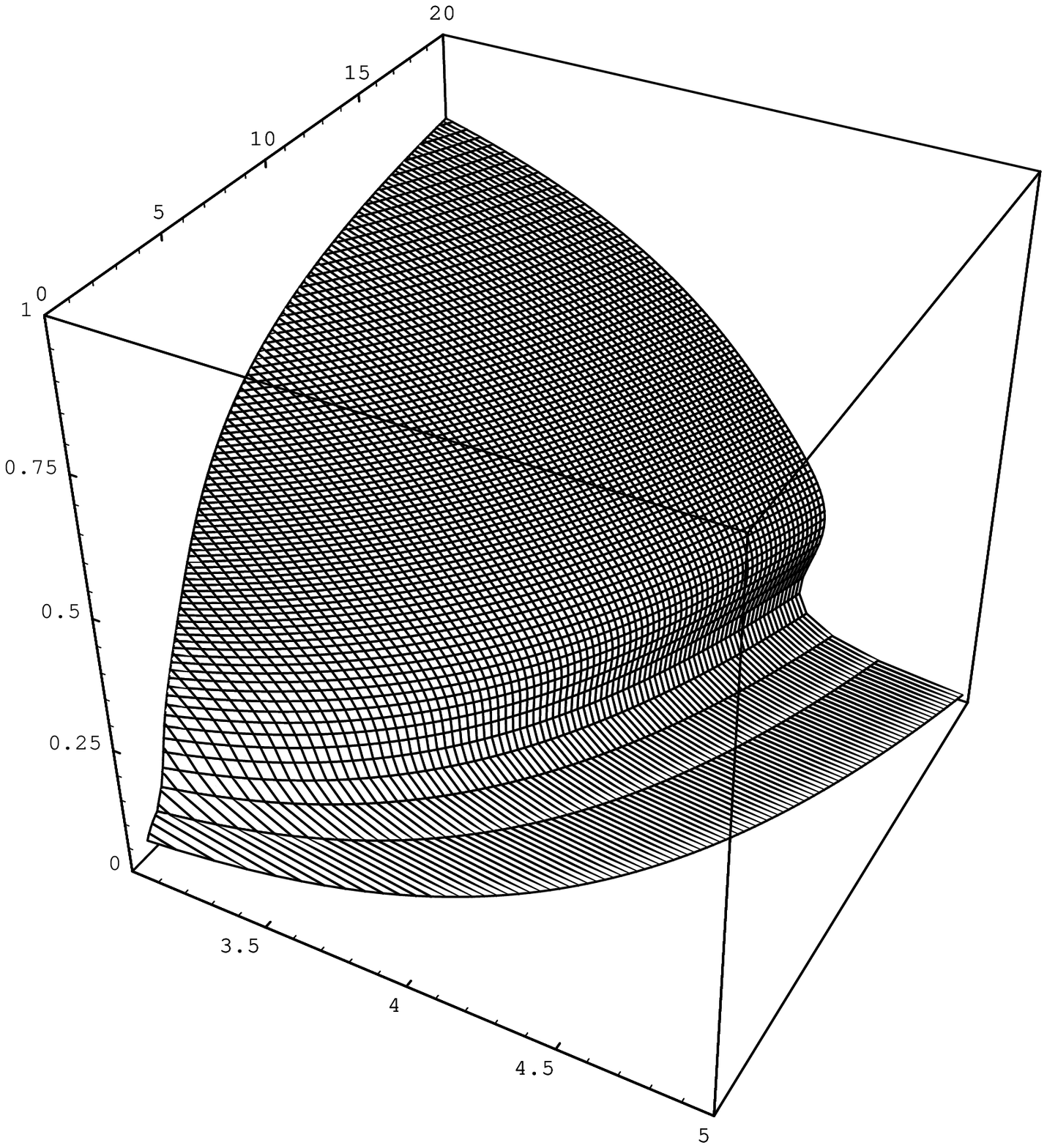}}}
\put(30,150){R}
\put(50,190){$\alpha$}
\put(60,110){$\eta$}
\put(0,0){\makebox(150,100)
          {\includegraphics{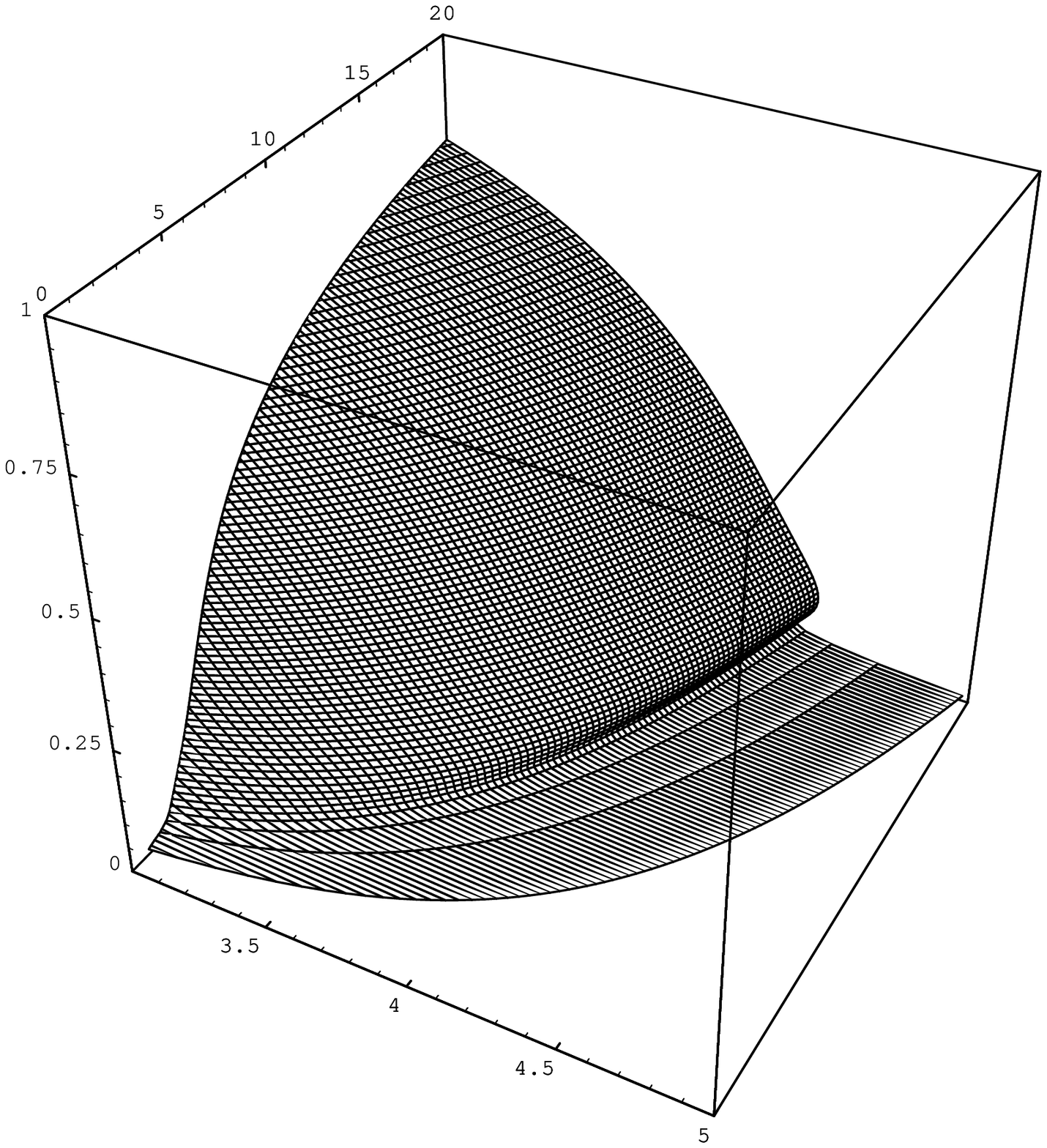}}}
\put(30,50){Q}
\put(50,90){$\alpha$}
\put(60,10){$\eta$}
\end{picture}
\end{center}
\caption{ 
Order parameters $R(\alpha,\eta)$ (top) and $Q(\alpha,\eta)$ (bottom) for MAP. 
As in fig.~2 and subsequent plots, we set $\gam=\gams=10$.
 }
\end{figure}
\begin{figure}[ht]
\begin{center}
\setlength{\unitlength}{1mm}
\begin{picture}(150,100)
\put(0,0){\makebox(150,100)
          {\includegraphics{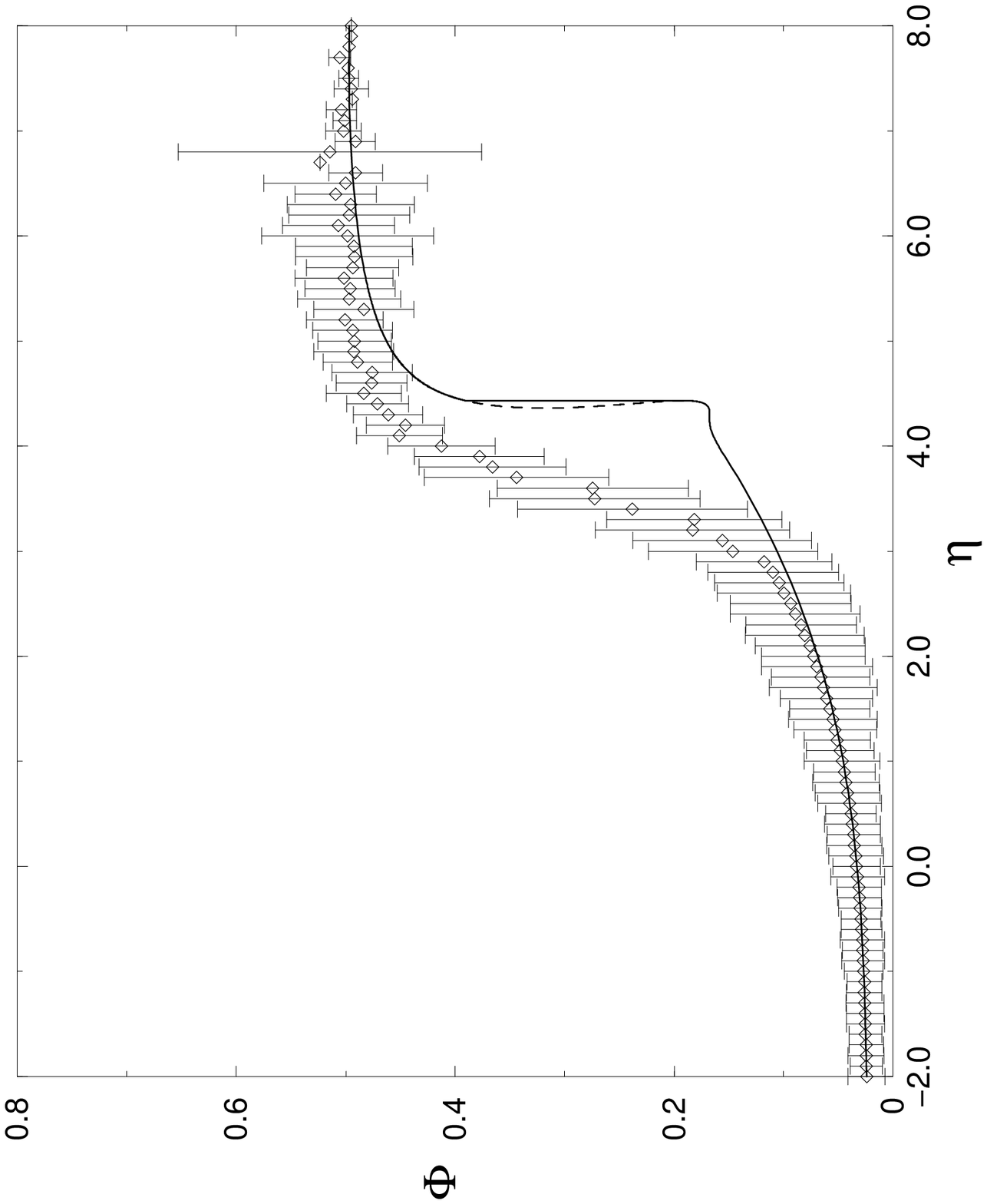}}}
\put(2,0){\makebox(150,100)
          {\includegraphics{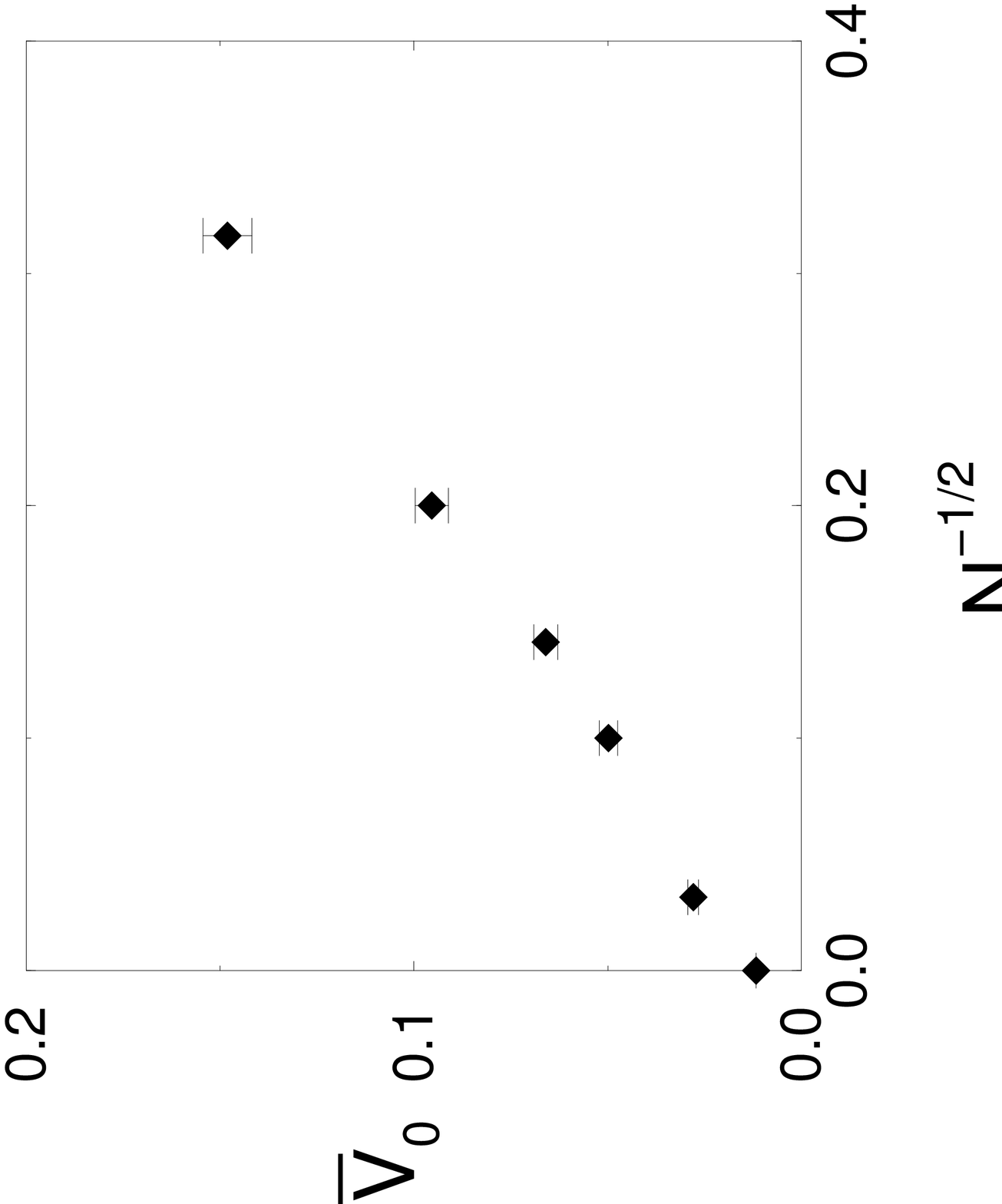}}}
\end{picture}
\end{center}
\caption{
Error $\Phi$ for soft selection
versus amount of outliers, represented by $\eta$. The relative
number of data is fixed at $\alpha=20$. The dashed part of the theoretical
curve denotes the region where three solutions of the saddlepoint equations
exist. The solid line follows the solution with minimal free energy.
Simulations are results from 100 runs with $N=100$.
Note that, for finite $N$, the transitions of the two orderparameters do not
coincide. The error measure $\Phi$ follows roughly the overlap $R$ between
solution vector and structure axis, whereas the drop in $Q$ gives rise to
the increased standard deviation at $\eta=6.8$.
The inset shows a finite size scaling of the phase transition as described 
in the text. The corresponding dimensions of the data are
$ N=10, 25, 50, 100, 1000$ respectively.
 }
\end{figure}
\begin{figure}[ht]
\begin{center}
\setlength{\unitlength}{1mm}
\begin{picture}(150,100)
\put(0,0){\makebox(150,100)
          {\includegraphics{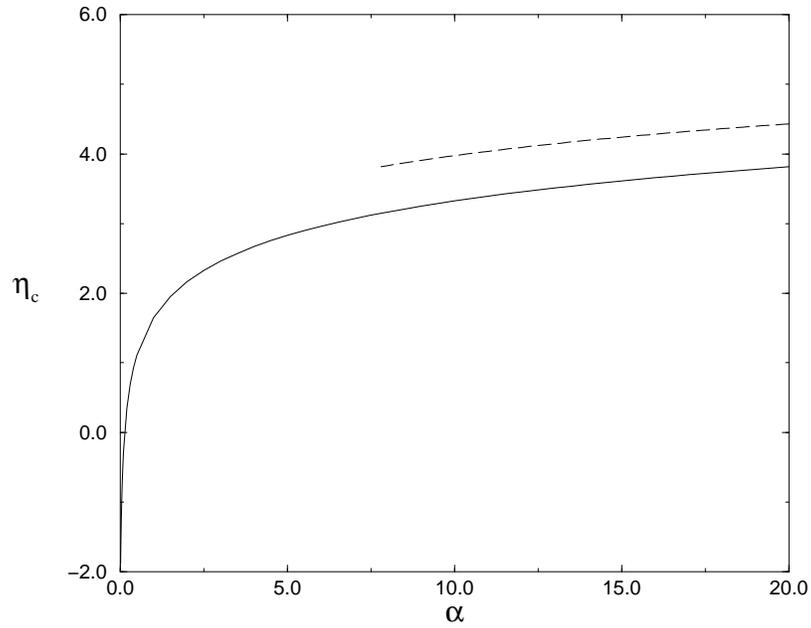}}}
\end{picture}
\end{center}
\caption{ 
Critical fraction of outliers for hard selection as a function of $\alpha$ 
(solid line).
The dashed line represents the phase transition for soft selection. 
 }
\end{figure}
\begin{figure}[ht]
\begin{center}
\setlength{\unitlength}{1mm}
\begin{picture}(150,100)
\put(0,0){\makebox(150,100)
          {\includegraphics{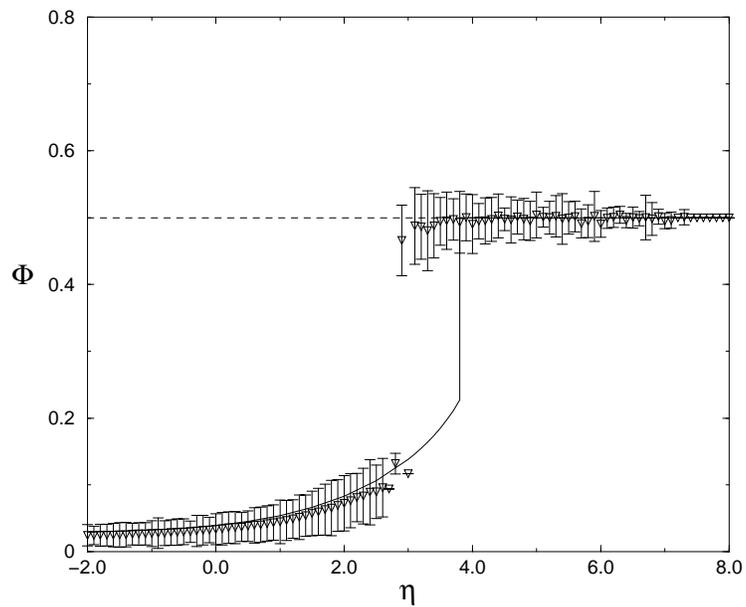}}}
\end{picture}
\end{center}
\caption{
Error $\Phi$ for hard selection versus amount of outliers, 
represented by $\eta$. As in Fig.~4, $\alpha=20$ and 
simulations are performed with $N=100$ and 100 runs.
The solid line indicates the theoretical result for the global optimum,
the dashed line for the local one.
 }
\end{figure}
\begin{figure}[ht]
\begin{center}
\setlength{\unitlength}{1mm}
\begin{picture}(150,100)
\put(0,0){\makebox(150,100)
          {\includegraphics{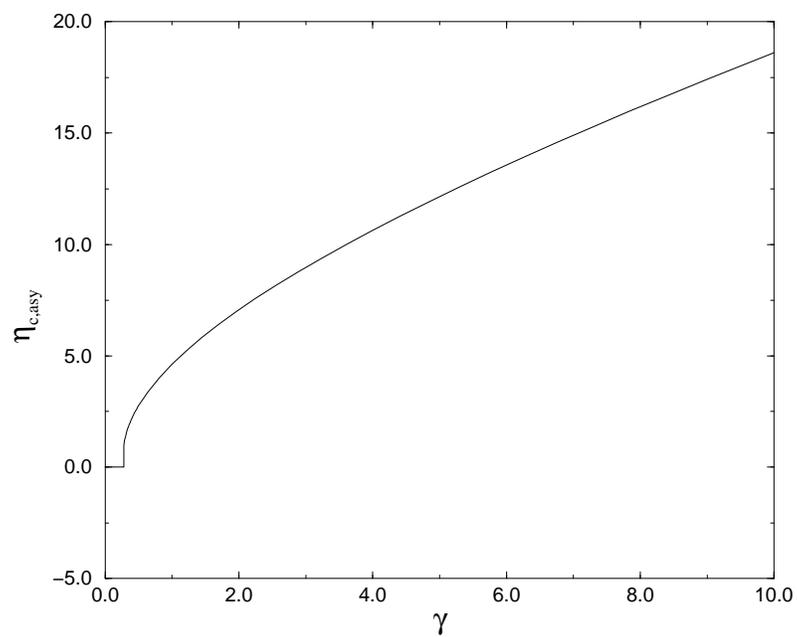}}}
\end{picture}
\end{center}
\caption{ 
Asymptotic critical fraction of outliers for hard selection, plotted against 
inverse squared width 
of the gaussian clusters.  
 }
\end{figure}

\end{document}